\documentclass[twocolumn]{autart}    

\usepackage{dsfont} 
\usepackage{graphicx} 
\usepackage{booktabs} 
\usepackage{amsmath} 

\newcommand{\ve}[1]{{\boldsymbol{#1}}} 
\newcommand{\tr}{\mbox{tr}} 
\newcommand{\ex}{\mathds{E}} 

\begin{document}
	
	\begin{frontmatter}
		
\title{Optimal controller/observer gains of discounted-cost LQG systems} 

\author[Delft]{Hildo~Bijl}\ead{h.j.bijl@tudelft.nl},    
\author[Uppsala]{Thomas~B.~Sch\"on}\ead{thomas.schon@it.uu.se}        

\address[Delft]{Delft Center for Systems and Control, Delft University of Technology, The Netherlands}  
\address[Uppsala]{Department of Information Technology, Uppsala University, Sweden}

\begin{keyword}                           
	Linear systems, cost function, LQG, optimal control, Riccati equation.               
\end{keyword}                             

\begin{abstract}                          
	The linear-quadratic-Gaussian (LQG) control paradigm is well-known in literature. The strategy of minimizing the cost function is available, both for the case where the state is known and where it is estimated through an observer. The situation is different when the cost function has an exponential discount factor, also known as a prescribed degree of stability. In this case, the optimal control strategy is only available when the state is known. This paper builds on from that result, deriving an optimal control strategy when working with an estimated state. Expressions for the resulting optimal expected cost are also given.
\end{abstract}
		
	\end{frontmatter}
	
	\section{Introduction}
	Consider the continuous-time linear system\footnote{From a formal point of view the system notation of~\eqref{eq:SystemEquationLQGFull} is incorrect, because $\ve{v}(t)$ and $\ve{w}(t)$ are not measurable with nonzero
		probability. However, since this notation is common in the control literature, we will stick with it. For methods to properly deal with stochastic differential equations, see~\cite{StochasticDEBook}.}
	\begin{subequations}
		\label{eq:SystemEquationLQGFull}
		\begin{align}
		\ve{\dot{x}}(t) & = A\ve{x}(t) + B\ve{u}(t) + \ve{v}(t),\\
		\ve{y}(t) & = C\ve{x}(t) + D\ve{u}(t) + \ve{w}(t),
		\end{align}
	\end{subequations}
	with $\ve{x}$ the state, $\ve{u}$ the input, $\ve{y}$ the output, $\ve{v}$ and $\ve{w}$ Gaussian white noise with respective intensities $V$ and $W$, and $A$, $B$, $C$ and $D$ the system matrices. We assume that the initial state $\ve{x_0}$ is unknown but distributed according to a Gaussian with  $\ve{\mu_0} = \ex[\ve{x_0}]$ and $\Sigma_0 = \ex\left[\ve{x_0}\ve{x_0}^T\right]$. Note that $\Sigma_0$ is \emph{not} the variance of $\ve{x_0}$.
	
	Our goal is to control system~\eqref{eq:SystemEquationLQGFull} such as to minimize the discounted (exponential) quadratic cost function
	\begin{equation}\label{eq:LQGCost}
	J(T) = \ex\left[\int_0^T e^{2\alpha t} \left(\ve{x}^T(t) Q \ve{x}(t) + \ve{u}^T(t) R \ve{u}(t)\right) \, dt\right],
	\end{equation}
	with $J(T)$ the expected cost, the real number $\alpha$ the discount exponent/prescribed degree of stability, and $Q \geq 0$ and $R > 0$ symmetric weight matrices. In particular, we will optimize the infinite-time expected cost $J$, with
	\begin{equation}\label{eq:LQGCostInfiniteTime}
	J = \lim_{T \rightarrow \infty} J(T).
	\end{equation}
	Our contribution in this paper is that we derive the optimal controller and observer gains for the continuous-time linear system~\eqref{eq:SystemEquationLQGFull} such that the expected cost $J$ given in~\eqref{eq:LQGCostInfiniteTime} is minimized.
	
	\section{Related work}\label{s:LQGOCLiterature}
	
	Linear-Quadratic-Gaussian (LQG) systems---linear systems with a quadratic cost function subject to Gaussian noise---have been thoroughly investigated in the past. This was especially true near the 1960s, with for instance the publication of the Kalman filter~\cite{KalmanFilterContinuousTime,KalmanFilter}.
	
	The discoveries from the decades afterwards have been summarized in numerous textbooks. Examples include the books by~\cite[Chapter 7]{StochasticControlBook}, \cite[Chapter 5]{LinearOptimalControlSystems}, \cite[Chapter 1]{OptimalControlBook2}, \cite[Chapters 3, 8]{LQBook}, \cite[Chapter 6]{OptimalControlBook}, \cite[Chapter 10]{LinearControlBook}, \cite[Chapter 9]{MFCBook} and \cite[Chapter 4]{DMCSBook}. All these books examine the non-discounted cost function (with $\alpha = 0$), save for~\cite[Section 3.5]{LQBook} that also considers  the discounted cost function, presenting results from an earlier paper~\cite{PrescribedDegreeOfStability}. Here it was shown that discounting the cost function is equivalent to prescribing a degree of stability.
	
	The prescribed degree of stability is actually a relevant problem in that it is a generalization of the regular LQG paradigm with the non-discounted cost function. There is also a variety of applications of this idea, such as fault tolerant flight control~\cite{LQGFaultTolerantControl}, spacecraft guidance~\cite{LQGSpacecraftGuidance} and robot manipulators~\cite{LQGRobotManipulator}. However, to the best of the authors' knowledge there are still fundamental properties remaining to be established and our contribution in this paper is to provide one of those. The work~\cite{PrescribedDegreeOfStability} examined  the situation where the state is assumed to be known. If the state can only be observed through noisy output measurements---a familiar problem for the non-discounted cost function---then we are not aware of any work that jointly optimize the controller and the state estimator. The closest is the work by~\cite{LQGDiscountedObserver}, who strived to derive a state estimator with minimal mean squared error, given a prescribed convergence rate. However, that work ignored the uncertainty in the initial state and did not examine the problem of jointly optimizing the controller and observer gains. In fact, it was not mentioned whether the separation principle still holds or not when using the discounted cost function. Hence, the problem of jointly optimizing the controller and observer gains, subject to a discounted cost function and an uncertain initial state, appears to be an open problem.

\section{Brief summary of known theorems}\label{s:SummaryOfKnownTheorems}

\noindent To place our new result in perspective, we briefly examine some known results first. We start with the non-discounted case ($\alpha = 0$) where the state $\ve{x}(t)$ is known (i.e., $C = I$ and $W = 0$). In this case the optimal control law is given by the following theorem.

\begin{thm}\label{th:OptimalControlNondiscountedFullState}
	Consider system~\eqref{eq:SystemEquationLQGFull}, where the state is assumed known. If $(A,B)$ is stabilizable, then the optimal control law minimizing the expected non-discounted cost $J$ (i.e., with $\alpha = 0$) is a linear control law $\ve{u}(t) = -F\ve{x}(t)$, where
	\begin{equation}
	F = R^{-1} B^T X,\label{eq:OptimalControlGain}
	\end{equation}
	and $X$ is the solution to the Riccati equation
	\begin{equation}\label{eq:RiccatiEquationForOptimalControl}
	A^T X + X A + Q - X B R^{-1} B^T X = 0.
	\end{equation}
	When $V = 0$, the corresponding expected cost equals
	\begin{equation}
	J = \ex\left[\ve{x_0}^T X \ve{x_0}\right] = \tr\left(X\Sigma_0\right).
	\end{equation}
	When $V \neq 0$, then $J(T) \rightarrow \infty$, but the steady-state cost rate equals
	\begin{equation}
	\lim_{T \rightarrow \infty} \frac{dJ(T)}{dT} = \tr\left(XV\right).\label{eq:SteadyStateCostRate}
	\end{equation}
\end{thm}
\begin{pf}
	See any of the aforementioned books; for example~\cite[Theorem 3.9]{LinearOptimalControlSystems}.
\end{pf}

There is another way to look at the Theorem~\ref{th:OptimalControlNondiscountedFullState}, which will become important in the proof of our main result. We know from~\cite[Theorem 3]{LQGPaper} that, for the above situation, and for any feedback matrix $F$, the expected steady-state cost rate equals
\begin{equation}\label{eq:appSteadyStateCostRate}
\lim_{T \rightarrow \infty} \frac{dJ(T)}{dT} = \tr\left(XV\right),
\end{equation}
where $X$ per definition is the unique solution to the Lyapunov equation
\begin{equation}\label{eq:appLyapunovEquation}
(A-BF)^T X + X (A-BF) + Q + F^T R F = 0.
\end{equation}
To minimize the above cost rate, we must find the value of $F$ minimizing~\eqref{eq:appSteadyStateCostRate}. Theorem~\ref{th:OptimalControlNondiscountedFullState} tells us that the cost rate~\eqref{eq:appSteadyStateCostRate} is minimized when $X$ satisfies~\eqref{eq:RiccatiEquationForOptimalControl} and $F$ subsequently equals~\eqref{eq:OptimalControlGain}. This is irrespective of the value of the positive definite matrix $V$.

Next, consider the case where there is a discount exponent $\alpha \neq 0$. Now the solution is given by the following Theorem. Note that $\alpha$ can be positive (a prescribed degree of stability) or negative (a discount exponent), but for ease of writing we always call it a discount exponent.

\begin{thm}\label{th:OptimalControlDiscountedFullState}
	Consider system~\eqref{eq:SystemEquationLQGFull}, where the state is assumed known. Define $A_\alpha = A + \alpha I$. If $(A_\alpha,B)$ is stabilizable, then the optimal control law minimizing the expected discounted cost $J$ is a linear control law $\ve{u}(t) = -F_\alpha \ve{x}(t)$, where
	\begin{equation}\label{eq:OptimalControlGainDiscounted}
	F_\alpha = R^{-1} B^T X_\alpha,
	\end{equation}
	and $X_\alpha$ is the solution to the Riccati equation
	\begin{equation}\label{eq:RiccatiEquationDiscounted}
	A_\alpha^T X_\alpha + X_\alpha A_\alpha + Q - X_\alpha B R^{-1} B^T X_\alpha = 0.
	\end{equation}
	The corresponding expected cost (for both zero and nonzero $V$) when $\alpha < 0$ equals
	\begin{equation}\label{eq:ExpectedDiscountedCostFullState}
	J = \tr\left(X_\alpha \left(\Sigma_0 - \frac{V}{2\alpha}\right)\right).
	\end{equation}
	When $\alpha \geq 0$, then $J(T) \rightarrow \infty$ as $T \rightarrow \infty$.
\end{thm}
\begin{pf}
	A proof is given by~\cite[Section 3.5]{LQBook}.
\end{pf}

	When the state is unknown, an observer needs to be used. The state estimate $\ve{\hat{x}}$ of this observer is updated through
	\begin{equation}\label{eq:ObserverForLQG}
	\ve{\dot{\hat{x}}}(t) = A\ve{\hat{x}}(t) + B\ve{u}(t) + K \left(\ve{y}(t) \hspace{-1pt} - \hspace{-1pt} C\ve{\hat{x}}(t) \hspace{-1pt} - \hspace{-1pt} D\ve{u}(t)\right), \hspace{-2pt}
	\end{equation}
	subject to some initial state estimate $\ve{\hat{x}}(0)$. If the state estimation error $\ve{e}(t)$ is defined as $\ve{e}(t) = \ve{\hat{x}}(t) - \ve{x}(t)$, then this error (i.e., its variance) can be minimized through the following Theorem.
	
	\begin{thm}\label{th:OptimalObserverNondiscounted}
		Consider system~\eqref{eq:SystemEquationLQGFull}. If $(A,C)$ is detectable, then the optimal observer gain minimizing the steady-state error covariance is
		\begin{equation}\label{eq:OptimalObserverGain}
		K = E C^T W^{-1},
		\end{equation}
		where $E$ is the optimal steady-state error covariance, found through
		\begin{equation}\label{eq:RiccatiEquationForOptimalObserver}
		AE + EA^T + V - E C^T W^{-1} C E = 0.
		\end{equation}
	\end{thm}
	\begin{pf}
		This is the famous Kalman-Bucy filter from~\cite{KalmanFilterContinuousTime}. A proof can also be found in~\cite[Theorem 4.5]{LinearOptimalControlSystems}.
	\end{pf}
	
	The above result holds regardless of the value of $\alpha$, because it is unrelated to the cost $J$. If our goal is to optimize the cost $J$ subject to $\alpha = 0$ (the non-discounted case) then the following Theorem provides the solution.
	
	\begin{thm}\label{th:OptimalControlNondiscounted}
		Consider system~\eqref{eq:SystemEquationLQGFull}. If $(A,B)$ is stabilizable and $(A,C)$ is detectable, then the optimal control law minimizing the expected non-discounted cost (i.e., with $\alpha = 0$) is a linear control law $\ve{u}(t) = -F\ve{\hat{x}}(t)$, with $F$ given by~\eqref{eq:OptimalControlGain}, $\ve{\hat{x}}(t)$ following from~\eqref{eq:ObserverForLQG} and the observer gain $K$ taken as~\eqref{eq:OptimalObserverGain}. The resulting expected steady-state cost rate is given by
		\begin{align}\label{eq:ExpectedSteadyStateCostRateForOptimalControl}
		\lim_{T \rightarrow \infty} \frac{dJ(T)}{dT} & = \tr\left(XKWK^T + EQ\right) \nonumber \\
		& = \tr\left(XV + EF^TRF\right),
		\end{align}
		with $X$ the solution of~\eqref{eq:RiccatiEquationForOptimalControl} and $E$ the solution of~\eqref{eq:RiccatiEquationForOptimalObserver}.
	\end{thm}
	\begin{pf}
		The optimal controller and observer gains follow from the separation principle. See for instance~\cite[Theorem 5.4]{LinearOptimalControlSystems}. Expressions for the expected steady-state cost rate can be derived using~\cite[Theorem 3]{LQGPaper}.
	\end{pf}
	
\section{Optimizing the discounted cost function}\label{s:OptimizingTheDiscountedCost}

In this section we derive the main result: the optimal controller/observer gains minimizing the discounted cost function, subject to an unknown state. It is important to realize that `optimal' here only means that the expected discounted cost~\eqref{eq:LQGCost} is minimized. There is no guarantee that the steady-state error variance, or any other quantity, is still at a minimum.

\begin{thm}\label{th:OptimalControlDiscounted}
	Consider system~\eqref{eq:SystemEquationLQGFull}. If $(A_\alpha,B)$ is stabilizable and $(A_\alpha,C)$ is detectable, then the optimal control law minimizing the expected discounted cost $J$ is a linear control law $\ve{u}(t) = -F_\alpha \ve{\hat{x}}(t)$, with $F_\alpha$ given by~\eqref{eq:OptimalControlGainDiscounted} and $X_\alpha$ given by~\eqref{eq:RiccatiEquationDiscounted}. Identically to~\eqref{eq:ObserverForLQG}, $\ve{\hat{x}}(t)$ is provided by the observer
	\begin{equation}\label{eq:OberverForDiscountedLQG}
	\hspace{0pt} \ve{\dot{\hat{x}}}(t) = A\ve{\hat{x}}(t) + B\ve{u}(t) + K_\alpha \left(\ve{y}(t) \hspace{-1pt} - \hspace{-1pt} C\ve{\hat{x}}(t) \hspace{-1pt} - \hspace{-1pt} D\ve{u}(t)\right), \hspace{-6pt}
	\end{equation}
	where $\ve{\hat{x}_0}$ is set to $\ve{\mu_0}$, the observer gain $K_\alpha$ is given by
	\begin{equation}\label{eq:OptimalObserverGainDiscounted}
	K_\alpha = E_\alpha C^T W^{-1}
	\end{equation}
	and $E_\alpha$ is the solution to the Riccati equation
	\begin{align}\label{eq:RiccatiEquationOfDiscountedObserver}
	A_\alpha E_\alpha + E_\alpha A_\alpha^T + \left(V - 2\alpha \left(\Sigma_0 - \ve{\mu_0}\ve{\mu_0}^T\right)\right) \hspace{7pt} & \nonumber \\
	- E_\alpha C^T W^{-1} C E_\alpha & = 0.
	\end{align}
	The corresponding expected cost for $\alpha < 0$ equals
	\begin{align}\label{eq:ExpectedDiscountedCostForOptimalControl}
	\hspace{-8pt}J & = \frac{1}{-2\alpha} \hspace{2pt} \tr\left(X_\alpha K_\alpha W K_\alpha^T + E_\alpha Q\right) + \ve{\mu_0}^T X_\alpha \ve{\mu_0} \nonumber \\
	& = \frac{1}{-2\alpha} \hspace{2pt} \tr\left(X_\alpha V + E_\alpha F_\alpha^T R F_\alpha\right) + \tr\left(X_\alpha \Sigma_0\right).
	\end{align}
	When $\alpha \geq 0$, then $J(T) \rightarrow \infty$ as $T \rightarrow \infty$.
\end{thm}
\begin{pf}
	To start, we write the joint dynamics of the system and its observer as
	\begin{align}
	\hspace{-8pt} \begin{bmatrix}
	\ve{\dot{x}}(t) \\
	\ve{\dot{e}}(t)
	\end{bmatrix} & = \begin{bmatrix}
	A - BF_\alpha & -B F_\alpha \\
	0 & A - K_\alpha C
	\end{bmatrix} \begin{bmatrix}
	\ve{x}(t) \\
	\ve{e}(t)
	\end{bmatrix} + \begin{bmatrix}
	\ve{v}(t) \\
	K_\alpha \ve{w}(t) - \ve{v}(t)
	\end{bmatrix} \nonumber \\
	& = \tilde{A} \ve{\tilde{x}}(t) + \ve{\tilde{v}}(t), \label{eq:LQGAdjustedSystem}
	\end{align}
	and the total expected cost as 
	\begin{equation}
	J(T) = \ex\left[\int_0^T e^{2\alpha t} \ve{\tilde{x}}^T(t) \tilde{Q} \ve{\tilde{x}}(t) \, dt\right].
	\end{equation}
	Note that the tilde-notation used above denotes properties of the joint dynamics. We have already defined $\tilde{A}$, $\ve{\tilde{x}}(t)$ and $\ve{\tilde{v}}(t)$ as above. The variance $\tilde{V}$ of $\ve{\tilde{v}}$, the mean and variance of $\ve{\hat{x}_0}$ and the weight matrix $\tilde{Q}$ satisfy
\begin{subequations}
	\begin{align}
	\hspace{-6pt} \tilde{V} & = \begin{bmatrix}
	V & -V \\
	-V & K_\alpha WK_\alpha^T + V
	\end{bmatrix}, \\
	\hspace{-6pt} \ve{\tilde{\mu}_0} & = \ex\left[\ve{\tilde{x}_0}\right] = \ex\begin{bmatrix}
	\ve{x_0} \\
	\ve{e_0}
	\end{bmatrix} = \begin{bmatrix}
	\ve{\mu_0} \\
	\ve{0}
	\end{bmatrix}, \\
	\hspace{-6pt} \tilde{\Sigma}_0 & = \ex\left[\ve{\tilde{x}_0} \ve{\tilde{x}_0}^T\right] = \begin{bmatrix}
	\ve{\mu_0}\ve{\mu_0}^T & \ve{\mu_0}\ve{\mu_0}^T - \Sigma_0 \\
	\ve{\mu_0}\ve{\mu_0}^T - \Sigma_0 & \Sigma_0 - \ve{\mu_0}\ve{\mu_0}^T
	\end{bmatrix}, \hspace{-6pt} \\
	\hspace{-6pt} \tilde{Q} & = \begin{bmatrix}
	Q + F_\alpha^T R F_\alpha & F_\alpha^T R F_\alpha \\
	F_\alpha^T R F_\alpha & F_\alpha^T R F_\alpha
	\end{bmatrix}.
	\end{align}
\end{subequations}	
	Our goal is to choose $F_\alpha$ and $K_\alpha$ so as to minimize the expected cost $J$. This cost, according to~\cite[Theorem 2]{LQGPaper}, equals
	\begin{equation}
	J = \tr\left(\tilde{X}_\alpha \left(\tilde{\Sigma}_0 - \frac{\tilde{V}}{2\alpha}\right)\right), \label{eq:ExpectedCostStartingPointForProof}
	\end{equation}
	where $\tilde{X}_\alpha$ per definition is the unique solution to
	\begin{equation}\label{eq:jointLyapunovEquation}
	\tilde{A}_\alpha^T \tilde{X}_\alpha + \tilde{X}_\alpha \tilde{A}_\alpha + \tilde{Q} = 0,
	\end{equation}
	and where $\tilde{A}_\alpha$ is defined as $\tilde{A} + \alpha I$. Expression~\eqref{eq:ExpectedCostStartingPointForProof} holds for any $F_\alpha$ and $K_\alpha$, which implies that we need to find the~$F_\alpha$ and~$K_\alpha$ that minimize it. Note that we cannot directly solve this by applying Theorem~\ref{th:OptimalControlNondiscountedFullState}, because this time $\tilde{V}$ is not constant: it depends on $K_\alpha$. We need a different method.
	
	First, we expand our matrix equations into elements. This turns~\eqref{eq:jointLyapunovEquation} into the following three equations,
	\begin{subequations}\label{jointLyapunovSubequations}
		\begin{align}
		\left(A_\alpha - BF_\alpha\right)^T \tilde{X}_\alpha^{11} + \tilde{X}_\alpha^{11} \left(A_\alpha - BF_\alpha\right) \hspace{8pt}  & \nonumber \\
		+ \left(Q + F_\alpha^T R F_\alpha\right) & = 0, \label{eq:jointLyapunovSubequations1} \\
		\left(A_\alpha - BF_\alpha\right)^T \tilde{X}_\alpha^{12} - \tilde{X}_\alpha^{11} BF_\alpha \hspace{20pt} & \nonumber \\
		+ \tilde{X}_\alpha^{12} \left(A_\alpha - K_\alpha C\right) + F_\alpha^T R F_\alpha & = 0, \label{eq:jointLyapunovSubequations2} \\
		\left(A_\alpha - K_\alpha C\right)^T \tilde{X}_\alpha^{22} - F_\alpha^T B^T \tilde{X}_\alpha^{12} \hspace{46pt} & \nonumber \\
		- \tilde{X}_\alpha^{21} BF_\alpha + \tilde{X}_\alpha^{22} \left(A_\alpha - K_\alpha C\right) + F_\alpha^T R F_\alpha & = 0. \label{eq:jointLyapunovSubequations3}
		\end{align}
	\end{subequations}
	There is a fourth expression, but it is identical to~\eqref{eq:jointLyapunovSubequations2} --- to be precise, it is its transpose --- so it is not worth mentioning. Similarly we can expand~\eqref{eq:ExpectedCostStartingPointForProof} as
	\begin{align}
	\hspace{-6pt} J & = \tr\left(\tilde{X}_\alpha^{11} \left(\Sigma_0 - \frac{V}{2\alpha}\right) - \tilde{X}_\alpha^{12} \left(\Sigma_0 - \ve{\mu_0}\ve{\mu_0}^T - \frac{V}{2\alpha}\right) \right. \nonumber \\
	& \hspace{16pt} - \tilde{X}_\alpha^{21} \left(\Sigma_0 - \ve{\mu_0}\ve{\mu_0}^T - \frac{V}{2\alpha}\right) \nonumber \\
	& \hspace{16pt} \left. + \tilde{X}_\alpha^{22} \left(\Sigma_0 - \ve{\mu_0}\ve{\mu_0}^T - \frac{V}{2\alpha} - \frac{K_\alpha W K_\alpha^T}{2\alpha}\right)\right). \hspace{-4pt} \label{eq:jointCostEquation}
	\end{align}
	It is difficult to jointly optimize $F_\alpha$ and $K_\alpha$ to minimize the above cost. The key here is to first assume a certain value for $F_\alpha$ and then find the value of $K_\alpha$ that is optimal for this particular value of $F_\alpha$. To be precise, we assume that $F_\alpha$ is given by~\eqref{eq:OptimalControlGainDiscounted}.
	
	It is interesting to note that this value for $F_\alpha$ happens to optimize the first term of~\eqref{eq:jointCostEquation},
	\begin{equation}\label{eq:ExpectedCostPart11}
	J_{11} = \tr\left(\tilde{X}_\alpha^{11} \left(\Sigma_0 - \frac{V}{2\alpha}\right)\right).
	\end{equation}
	After all, from~\eqref{eq:jointLyapunovSubequations1} we see that $\tilde{X}_\alpha^{11}$ solely depends on $F_\alpha$ and not on $K_\alpha$. The problem of optimizing $F_\alpha$ now turns out to be equivalent to the problem solved by  Theorem~\ref{th:OptimalControlDiscountedFullState}. (Also see the note after Theorem~\ref{th:OptimalControlNondiscountedFullState}.) It follows that the value of $F_\alpha$ minimizing $J_{11}$ equals~\eqref{eq:OptimalControlGainDiscounted}, and that $\tilde{X}_\alpha^{11}$ from~\eqref{eq:jointLyapunovSubequations1} equals the solution $X_\alpha$ of~\eqref{eq:RiccatiEquationDiscounted}.
	
	For this assumed value of $F_\alpha$, the other equations greatly simplify. If we insert~\eqref{eq:OptimalControlGainDiscounted} into~\eqref{eq:jointLyapunovSubequations2}, we directly find that $\tilde{X}_\alpha^{12} = 0$. This tells us that the separation principle still holds for this situation, albeit in an adjusted form. At the same time~\eqref{eq:jointLyapunovSubequations3} reduces to
	\begin{equation}
	\left(A_\alpha - K_\alpha C\right)^T \tilde{X}_\alpha^{22} + \tilde{X}_\alpha^{22} \left(A_\alpha - K_\alpha C\right) + F_\alpha^T R F_\alpha = 0.
	\end{equation}
	Our goal is to find the value of $K_\alpha$ minimizing the last term from~\eqref{eq:jointCostEquation}. That is, we want to minimize
	\begin{equation}
	J_{22} = \tr\left(\tilde{X}_\alpha^{22} \left(\Sigma_0 - \ve{\mu_0}\ve{\mu_0}^T - \frac{V}{2\alpha} - \frac{K_\alpha W K_\alpha^T}{2\alpha}\right)\right).
	\end{equation}
	According to~\cite[Theorem 16]{LQGPaper}, we can rewrite this as
	\begin{equation}\label{eq:ExpectedCostPart22}
	J_{22} = \tr\left(\left(-\frac{E_\alpha}{2\alpha}\right) F_\alpha^T R F_\alpha\right),
	\end{equation}
	where the term $\left(-\frac{E_\alpha}{2\alpha}\right)$ per definition satisfies
	\begin{align}
	\left(A_\alpha - K_\alpha C\right) \left(-\frac{E_\alpha}{2\alpha}\right) + \left(-\frac{E_\alpha}{2\alpha}\right) \left(A_\alpha - K_\alpha C\right)^T \hspace{-24pt} & \nonumber \\
	+ \left(\Sigma_0 - \ve{\mu_0}\ve{\mu_0}^T - \frac{V}{2\alpha}\right) - \frac{K_\alpha W K_\alpha^T}{2\alpha} & = 0.
	\end{align}
	From this, we can directly find the value of $K_\alpha$ minimizing~\eqref{eq:ExpectedCostPart22}, and hence minimizing the total expected cost~$J$. According to the principle described right after Theorem~\ref{th:OptimalControlNondiscountedFullState}, it equals~\eqref{eq:OptimalObserverGainDiscounted}.
	
	To summarize, we have assumed that $F_\alpha$ was given by~\eqref{eq:OptimalControlGainDiscounted} and subsequently found that the optimal value for $K_\alpha$ equals~\eqref{eq:OptimalObserverGainDiscounted}. Of course this does \emph{not} necessarily mean that this combination of $F_\alpha$ and $K_\alpha$ jointly optimizes the expected cost $J$. We need one more step.
	
	For this step, we have to reverse the process: first we assume that $K_\alpha$ equals~\eqref{eq:OptimalObserverGainDiscounted} and subsequently we optimize the cost $J$ for $F_\alpha$. When doing so, we do have to use a different system notation. Instead of considering the joint dynamics of $\ve{x}$ and $\ve{e}$ in~\eqref{eq:LQGAdjustedSystem}, we consider the joint dynamics of $\ve{\hat{x}}$ and $\ve{e}$. Additionally, instead of optimizing the cost $J$ written as~\eqref{eq:ExpectedCostStartingPointForProof}, we first use~\cite[Theorem 16]{LQGPaper} to rewrite the expression. If we follow these steps, then in an identical way we find that the optimal value of $F_\alpha$ equals~\eqref{eq:OptimalControlGainDiscounted}.
	
	To conclude, if we choose $F_\alpha$ as~\eqref{eq:OptimalControlGainDiscounted} then the optimal $K_\alpha$ equals~\eqref{eq:OptimalObserverGainDiscounted}, and vice versa if we choose $K_\alpha$ as~\eqref{eq:OptimalObserverGainDiscounted} then the optimal $F_\alpha$ equals~\eqref{eq:OptimalControlGainDiscounted}. This proves that our combination of $(F_\alpha, K_\alpha)$ is at least a local solution to the optimization problem. However, because the optimization problem is convex in both $F_\alpha$ and $K_\alpha$, it must also be the global optimum. Hence, we conclude that the combination of $F_\alpha$ and $K_\alpha$ minimizes the expected cost $J$.
	
	The only thing left to prove is the cost expression~\eqref{eq:ExpectedDiscountedCostForOptimalControl}. The second line from this equation follows directly from~\eqref{eq:ExpectedCostPart11} and~\eqref{eq:ExpectedCostPart22}: just add $J_{11}$ and $J_{22}$. The first line follows in the same way, if you redo the full derivation with the joint state of $\ve{\hat{x}}$ and $\ve{e}$, as described above. That concludes this proof.
\end{pf}

This theorem shows how to optimally trade off between compensating for process noise ($V$), for measurement noise ($W$) and for uncertainty in the initial state ($\Sigma_0 - \ve{\mu_0}\ve{\mu_0}^T$). None of the previously derived theorems had to include all these three parameters in their trade-off, which is what makes this new result significant.

Due to the separation principle, the stability of the controlled system is similar to when we applied Theorem~\ref{th:OptimalControlDiscountedFullState}. The eigenvalues of the closed-loop system are all guaranteed to be smaller than $-\alpha$~\cite[Section 3.5]{LQBook}. Hence, if $\alpha > 0$, stability is guaranteed.
	
	\section{Conclusions and recommendations}\label{s:LQGControlConclusion}
	
	Through Theorem~\ref{th:OptimalControlDiscounted} it is now possible to find the optimal controller and observer gains of an LQG system with discounted cost. This paper also serves as a focused overview of this part of control engineering.
	
	
	Future work on this subject can look into replacing the discount exponent $\alpha$ by a discount matrix, investigate the effect of a finite time window $T$ on the optimal controller/observer parameters, or examine time-varying systems, similarly to~\cite{LQGDiscountedObserver}, to see whether the same results are still applicable.

	\begin{ack}
		This research is supported by the Dutch Technology Foundation STW, which is part of the Netherlands Organisation for Scientific  Research (NWO), and which is partly funded by the Ministry of Economic Affairs (Project number: 12173, SMART-WIND). The work was also supported by the Swedish research Council (VR) via the project \emph{NewLEADS - New Directions in Learning Dynamical Systems} (Contract number: 621-2016-06079) and by the Swedish Foundation for Strategic Research (SSF) via the project \emph{ASSEMBLE} (Contract number: RIT15-0012).
	\end{ack}
	
	\bibliographystyle{plain}        
	\bibliography{bibliography}           

\end{document}